\documentclass[twocolumn, preprintnumbers,prl,aps,amssymb,showpacs,superscriptaddress]{revtex4}
\usepackage{graphicx}
\usepackage{natbib}
\usepackage{dcolumn}
\usepackage{wasysym}
\usepackage{bm} \usepackage{color}
\begin{document}

\newcommand{\ie}{{\it i.e.}}
\newcommand{\eg}{{\it e.g.}}
\newcommand{\etal}{{\it et al.}}
\newcommand{\K}{Ba$_{1-x}$K$_x$Fe$_2$As$_2$}
\newcommand{\KFeAs}{KFe$_2$As$_2$}
\newcommand{\Co}{Ba(Fe$_{1-x}$Co$_x$)$_2$As$_2$}
\newcommand{\Kzero}{$\kappa_0/T$}
\newcommand{\Tc}{$T_c \cong$}
\newcommand{\units}{$\mu \text{W}/\text{K}^2\text{cm}$}
\newcommand{\p}[1]{\left( #1 \right)}
\newcommand{\Dd}[2]{\frac{\text{d} #1}{\text{d}#2}}


\title{
Universal Heat Conduction in the Iron-Arsenide Superconductor KFe$_2$As$_2$~:\\
Evidence of a $d$-wave State
}


\author{J.-Ph.~Reid}
\affiliation{D\'epartement de physique \& RQMP, Universit\'e de Sherbrooke, Sherbrooke, Qu\'ebec, Canada J1K 2R1}

\author{M.~A.~Tanatar}
\affiliation{Ames Laboratory, Ames, Iowa 50011, USA}

\author{A.~Juneau-Fecteau} 
\affiliation{D\'epartement de physique \& RQMP, Universit\'e de Sherbrooke, Sherbrooke, Qu\'ebec, Canada J1K 2R1}

\author{R.~T.~Gordon} 
\affiliation{D\'epartement de physique \& RQMP, Universit\'e de Sherbrooke, Sherbrooke, Qu\'ebec, Canada J1K 2R1}

\author{S.~Ren\'e~de~Cotret} 
\affiliation{D\'epartement de physique \& RQMP, Universit\'e de Sherbrooke, Sherbrooke, Qu\'ebec, Canada J1K 2R1}

\author{N.~Doiron-Leyraud} 
\affiliation{D\'epartement de physique \& RQMP, Universit\'e de Sherbrooke, Sherbrooke, Qu\'ebec, Canada J1K 2R1}

\author{T.~Saito} 
\affiliation{Chiba University \& JST-TRIP, Japan}

\author{H.~Fukazawa} 
\affiliation{Chiba University \& JST-TRIP, Japan}

\author{Y.~Kohori}
\affiliation{Chiba University \& JST-TRIP, Japan}

\author{K.~Kihou}
\affiliation{AIST \& JST-TRIP, Japan}

\author{ C.~H.~Lee} 
\affiliation{AIST \& JST-TRIP, Japan}

\author{A.~Iyo} 
\affiliation{AIST \& JST-TRIP, Japan}

\author{H.~Eisaki}
\affiliation{AIST \& JST-TRIP, Japan}

\author{R.~Prozorov}
\affiliation{Ames Laboratory, Ames, Iowa 50011, USA} 
\affiliation{Department of Physics and Astronomy, Iowa State University, Ames, Iowa 50011, USA }

\author{Louis Taillefer}
\altaffiliation{E-mail: louis.taillefer@physique.usherbrooke.ca }
\affiliation{D\'epartement de physique \& RQMP, Universit\'e de Sherbrooke, Sherbrooke, Qu\'ebec, Canada J1K 2R1}
\affiliation{Canadian Institute for Advanced Research, Toronto, Ontario, Canada M5G 1Z8}

\date{\today}


\begin{abstract}

The thermal conductivity $\kappa$ of the iron-arsenide superconductor \KFeAs\ was measured down to 50 mK
 for a heat current parallel and perpendicular to the tetragonal $c$ axis.
A residual linear term at $T \to 0$, $\kappa_0/T$, is observed for both current directions, 
confirming the presence of nodes in the superconducting gap.
Our value of $\kappa_0/T$ in the plane is equal to that reported by Dong {\it et al.}~[Phys.~Rev.~Lett.~{\bf 104}, 087005 (2010)] 
for a sample whose residual resistivity $\rho_0$ was ten times larger. 
This independence of $\kappa_0/T$ on impurity scattering is the signature of universal heat transport,
a property of superconducting states with symmetry-imposed line nodes.
This argues against an $s$-wave state with accidental nodes.
It favors instead a $d$-wave state, 
an assignment consistent with five additional properties:
the magnitude of the critical scattering rate $\Gamma_c$ for suppressing $T_c$ to zero;
the magnitude of $\kappa_0/T$, and
its dependence on current direction and on magnetic field;
the temperature dependence of $\kappa(T)$. 


\end{abstract}


\maketitle


The pairing mechanism in a superconductor is intimately related to the pairing symmetry, which in turn is related to the gap structure $\Delta({\bf k})$.
In a $d$-wave state with $d_{x^2 - y^2}$ symmetry, the order parameter changes sign with angle in the $x$-$y$ plane, forcing the gap to go to zero along diagonal 
directions ($\pm k_x$=$\pm k_y$). 
Those zeros (or nodes) in the gap are imposed by symmetry. 
The gap in states with $s$-wave symmetry will in general not have nodes, although accidental nodes can occur depending on the anisotropy 
of the pairing interaction.
In iron-based superconductors, the gap shows nodes in some materials, as in BaFe$_2$(As$_{1-x}$P$_x$)$_2$ 
~\cite{Hashimoto2010}
and Ba(Fe$_{1-x}$Ru$_x$)$_2$As$_2$~\cite{Qiu2011},
and not in others, as in Ba$_{1-x}$K$_x$Fe$_2$As$_2$~\cite{Luo2009,Reid2011} and
Ba(Fe$_{1-x}$Co$_x$)$_2$As$_2$~\cite{Tanatar2010,Reid2010} at optimal doping.

In KFe$_2$As$_2$, the end-member of the Ba$_{1-x}$K$_x$Fe$_2$As$_2$ series 
(with $x=1$),
the presence of nodes was detected by thermal conductivity~\cite{Dong2010}, 
penetration depth \cite{Hashimoto2010a} and NMR \cite{Fukazawa2009,Zhang2010}. 
The question is whether those nodes are imposed by symmetry or accidental.
Calculations differ in their predictions~\cite{Graser2009,Maiti2011a,Suzuki2011}. 
Some favor a $d$-wave state~\cite{Thomale2011}, others an $s$-wave state 
with accidental line nodes that run either parallel to the $c$ axis~\cite{Maiti2011} or perpendicular~\cite{Suzuki2011}.
One can distinguish a $d$-wave state from an extended $s$-wave state with accidental nodes
by looking at the effect of impurity scattering~\cite{Borkowski1994}.
Nodes are robust in the former,
but impurity scattering will eventually 
remove them in the latter, as it makes $\Delta({\bf k})$ less anisotropic.

In this Letter, we investigate the pairing symmetry of KFe$_2$As$_2$ using 
thermal conductivity, 
a bulk directional probe of the superconducting gap~\cite{Shakeripour2009}. 
All aspects of heat transport are found to be in agreement with theoretical expectation for a $d$-wave gap~\cite{Graf1996,Durst2000},
and inconsistent with accidental line nodes, whether vertical or horizontal.
%
Moreover, the critical scattering rate $\Gamma_c$ for suppressing $T_c$ to zero is of order $T_{c0}$,
as expected for $d$-wave, while it is 50 times $T_{c0}$ in optimally-doped 
BaFe$_2$As$_2$~\cite{Kirshenbaum2012}.


{\it Methods.--}
Single crystals of KFe$_2$As$_2$ were grown from self flux \cite{Kihou2010}.
Two samples were measured: 
one for currents along the $a$ axis, 
one for currents along the $c$ axis. 
Their superconducting temperature, defined by the point of zero resistance, is 
$T_c = 3.80 \pm 0.05$~K 
and $3.65 \pm 0.05$~K, respectively.
Since the contacts were soldered with a superconducting alloy, a small magnetic field of 0.05~T was applied to make the contacts normal and thus ensure 
good thermalization. 
For more information on sample geometry, contact technique and measurement protocol, see ref.~\cite{Reid2010}.  
%


\begin{figure}[t]
\centering
\includegraphics[width=8.5cm]{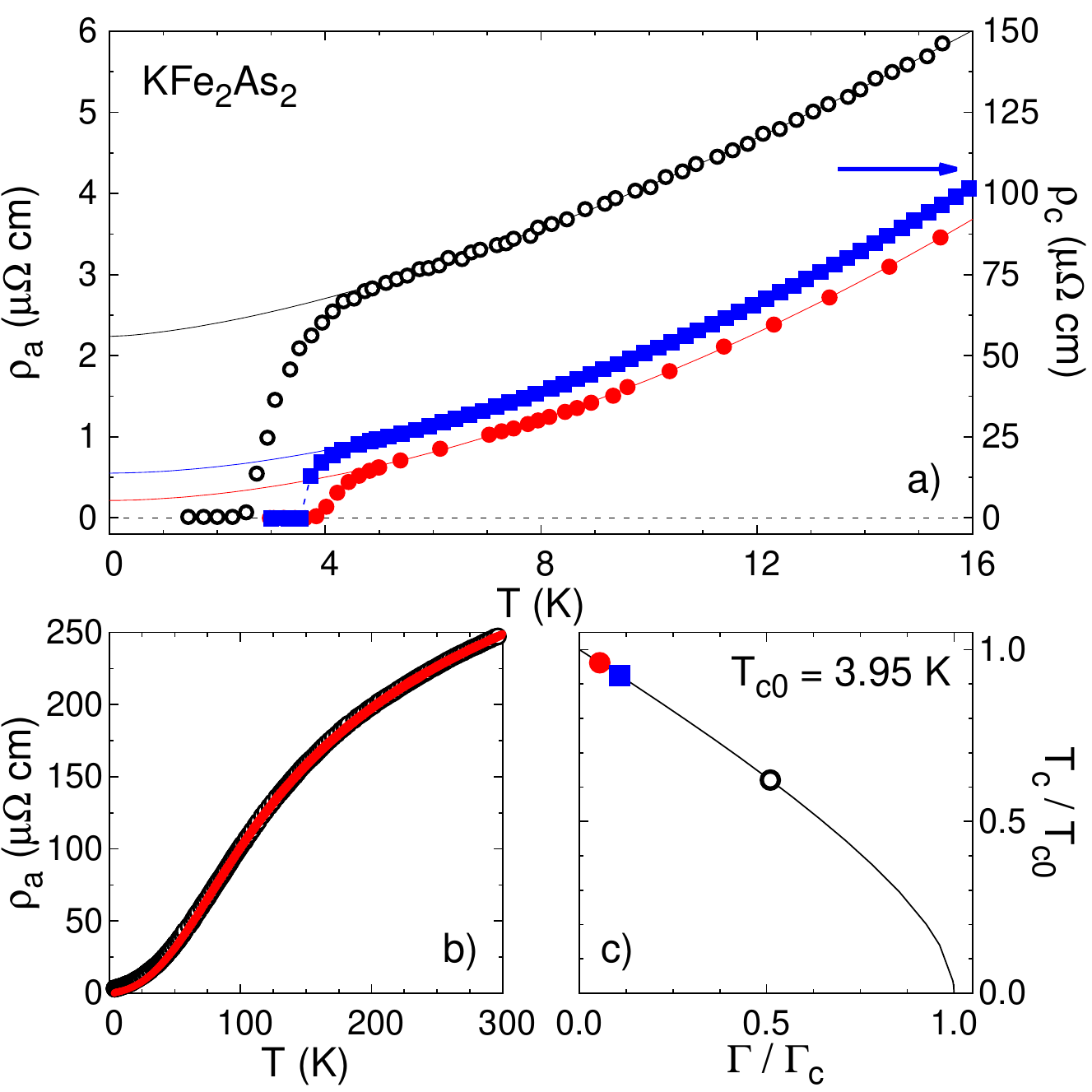}
\caption{
(a)
Electrical resistivity of the two samples of KFe$_2$As$_2$ studied here, with $J \parallel a$ (full red circles, left axis) 
and $J \parallel c$ (full blue squares, right axis).
Our $a$-axis data is compared to that of Dong {\it et al.}~\cite{Dong2010} (open circles, left axis), 
normalized here to have the same value at $T = 300$~K (see text). 
The lines are a fit to $\rho = \rho_0 + aT^\alpha$ from which  we extrapolate $\rho_0$ at $T=0$. 
(b)
Same data for the two $a$-axis samples, up to 300~K.
(c)
Abrikosov-Gorkov formula for the decrease of $T_c$ with scattering rate $\Gamma$ (line), used 
to obtain a value of  $\Gamma/\Gamma_c$ for the three samples of KFe$_2$As$_2$,
given their $T_c$ values and the factor 10 in $\rho_0$ between the two $a$-axis samples (circles), 
assuming a disorder-free value of $T_{c0}=3.95$~K. 
}
\label{Fig1}
\end{figure}



{\it Resistivity.--}
To study the effect of impurity scattering in KFe$_2$As$_2$, we performed measurements on a single crystal whose residual resistivity ratio
(RRR) is 10 times larger than that of the sample studied by Dong {\it et al.}~\cite{Dong2010} (Fig.~1a).
To remove the uncertainty associated with geometric factors, we normalize the data of Dong {\it et al.} to our value at $T = 300$~K.
A power-law fit below 16~K yields a residual resistivity 
$\rho_0 = 0.21 \pm 0.02~\mu \Omega$~cm 
($2.24 \pm 0.05~\mu \Omega$~cm)
for our (their) sample, so that $\rho(300~{\rm K}) / \rho_0$ = 1180 and 110, respectively.


\begin{figure}[t]
\centering
\includegraphics[width=8.5cm]{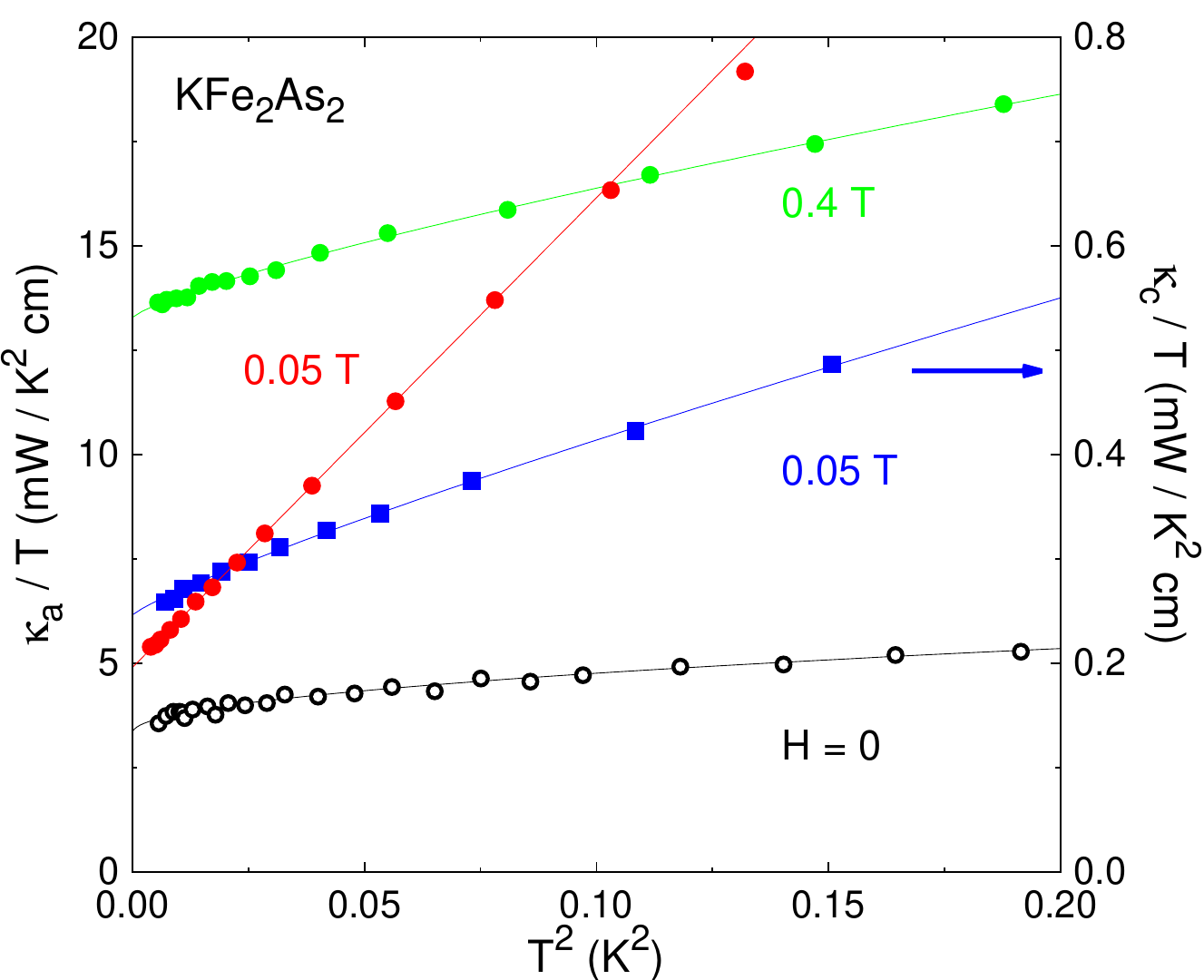}
\caption{
Thermal conductivity of KFe$_2$As$_2$, plotted as $\kappa/T$ vs $T^2$,  for $J\parallel a$ ($\kappa_a$, circles, left axis) 
and $J\parallel c$ ($\kappa_c$, squares, right axis), for a magnetic field 
$H \parallel c$ as indicated.
Our $a$-axis data is compared to that of Dong {\it et al.}~\cite{Dong2010} (open circles, left axis), 
normalized by the same factor as in Fig.~1 (see text). 
Lines are a fit to $\kappa/T = a + bT^\alpha$, used to extrapolate the residual linear term $a \equiv \kappa_0/T$ at $T=0$. 
For our $a$-axis sample (full red circles), $\alpha = 2.0$, while for others $\alpha < 2$.
}
\label{Fig2}
\end{figure}


We attribute the lower $\rho_0$ 
in our sample to a lower concentration of impurities or defects.
Note that except for the different $\rho_0$, the two resistivity curves $\rho(T)$ are essentially 
identical (Fig.~1b).
Supporting evidence for a difference in impurity/defect concentration is the difference in critical temperature:
$T_c = 3.80 \pm 0.05$~K ($2.45 \pm 0.10$~K) 
for our (their) sample.
Assuming that the impurity scattering rate $\Gamma \propto \rho_0$, we can use the Abrikosov-Gorkov formula for the drop in $T_c$ vs $\Gamma$
to extract a value of $\Gamma / \Gamma_c$ for the two samples, where $\Gamma_c$ is the critical scattering rate needed to suppress $T_c$ to zero
(Fig.~1c).
We get 
$\Gamma / \Gamma_c = 0.05$ (0.5)
 for our (their) sample.

The $c$-axis resistivity $\rho_c(T)$ has the same temperature dependence as $\rho_a(T)$ below $T \simeq 40$~K
(Fig.~1a),
with an intrinsic anisotropy 
$\Delta \rho_c / \Delta \rho_a = 25~\pm~1$, where $\Delta \rho \equiv \rho(T) - \rho_0$,
with $\rho_{c0} = 13 \pm 1~\mu \Omega$~cm.
We attribute the larger anisotropy at $T \to 0$, $\rho_{c0} / \rho_{a0} = 60 \pm 10$, to a larger $\Gamma$ in our $c$-axis sample,
consistent with the lower value of $T_c$, from which we deduce
$\Gamma / \Gamma_c = 0.1$
(Fig.~1c).


{\it Universal heat transport.--}
The thermal conductivity is shown in
Fig.~2.
The residual linear term $\kappa_0/T$ is obtained from a fit to $\kappa / T = a + bT^\alpha$ below 0.3~K, where $a \equiv \kappa_0/T$.
The dependence of $\kappa_0/T$ on magnetic field $H$ is shown in 
Fig.~3.
Extrapolation to $H=0$ yields 
%
$\kappa_{a0}/T = 3.6 \pm 0.5$~mW/K$^2$~cm
and
$\kappa_{c0}/T = 0.18 \pm 0.03$~mW/K$^2$~cm.
We compare to the data by Dong {\it et al}.~\cite{Dong2010}, 
normalized by the same factor as for electrical transport,
giving
$\kappa_{0a}/T = 3.32 \pm 0.03$~mW/K$^2$~cm.
At $H \to 0$, 
$\kappa_{a0}/T$ is the same in the two samples (inset of Fig.~3), 
within error bars.

This universal heat transport, whereby $\kappa_0/T$ is independent of the impurity scattering rate, is a classic signature of 
line nodes imposed by symmetry~\cite{Graf1996,Durst2000}.
Calculations show the residual linear term to be independent of scattering rate and phase shift~\cite{Graf1996}, 
and free of Fermi-liquid and vertex corrections~\cite{Durst2000}.
For a quasi-2D $d$-wave superconductor~\cite{Graf1996,Durst2000}:
\begin{equation}
\frac{\kappa_{0}}{T}~ \simeq ~\frac{\kappa_{00}}{T} ~\equiv ~\frac{\hbar}{2 \pi} ~ \frac{\gamma_{\rm N} v^2_{\rm F}}{\Delta_0}~~~~~~,
\end{equation}
where 
$\gamma_{\rm N}$ is the residual linear term in the normal-state electronic specific heat, 
$v_{\rm F}$ is the
Fermi velocity,
and 
the superconducting gap 
$\Delta = \Delta_0$cos($2 \phi$)~\cite{Graf1996a}.

ARPES measurements on KFe$_2$As$_2$ reveal a Fermi surface with three concentric hole-like cylinders centered on the $\Gamma$ point of the Brillouin zone,
labeled $\alpha$, $\beta$ and $\gamma$, and a 4th cylinder near the $X$ point~\cite{Sato2009,Yoshida2011}. 
dHvA measurements 
detect all of these surfaces except the $\beta$, and obtain Fermi velocities in reasonable agreement with ARPES dispersions, 
with an average value of $v_{\rm F} \simeq 4 \times 10^6$~cm/s~\cite{Terashima2010a}.
The measured effective masses 
account for approximately 80\% of the measured $\gamma_{\rm N} = 85 \pm 10$~mJ/K$^2$~mol~\cite{Abdel-Hafiez2011,Fukazawa2011}.
In $d$-wave symmetry, the gap in KFe$_2$As$_2$ will necessarily have nodes on 
all $\Gamma$-centered
Fermi surfaces, and possibly on the $X$-centered surface as well~\cite{Thomale2011}.
The total $\kappa_0/T$ may be estimated from Eq.~1 by using the average $v_{\rm F}$ and the measured (total) $\gamma_{\rm N}$,
which yields 
$\kappa_{00}/T = 3.3 \pm 0.5$~mW/K$^2$~cm, 
assuming $\Delta_0 = 2.14~k_{\rm B} T_{c0}$, with $T_{c0} = 3.95$~K.
This is in excellent agreement with the experimental value of
$\kappa_0/T = 3.6 \pm 0.5$~mW/K$^2$~cm.


\begin{figure}[t]
\centering
\includegraphics[width=8.5cm]{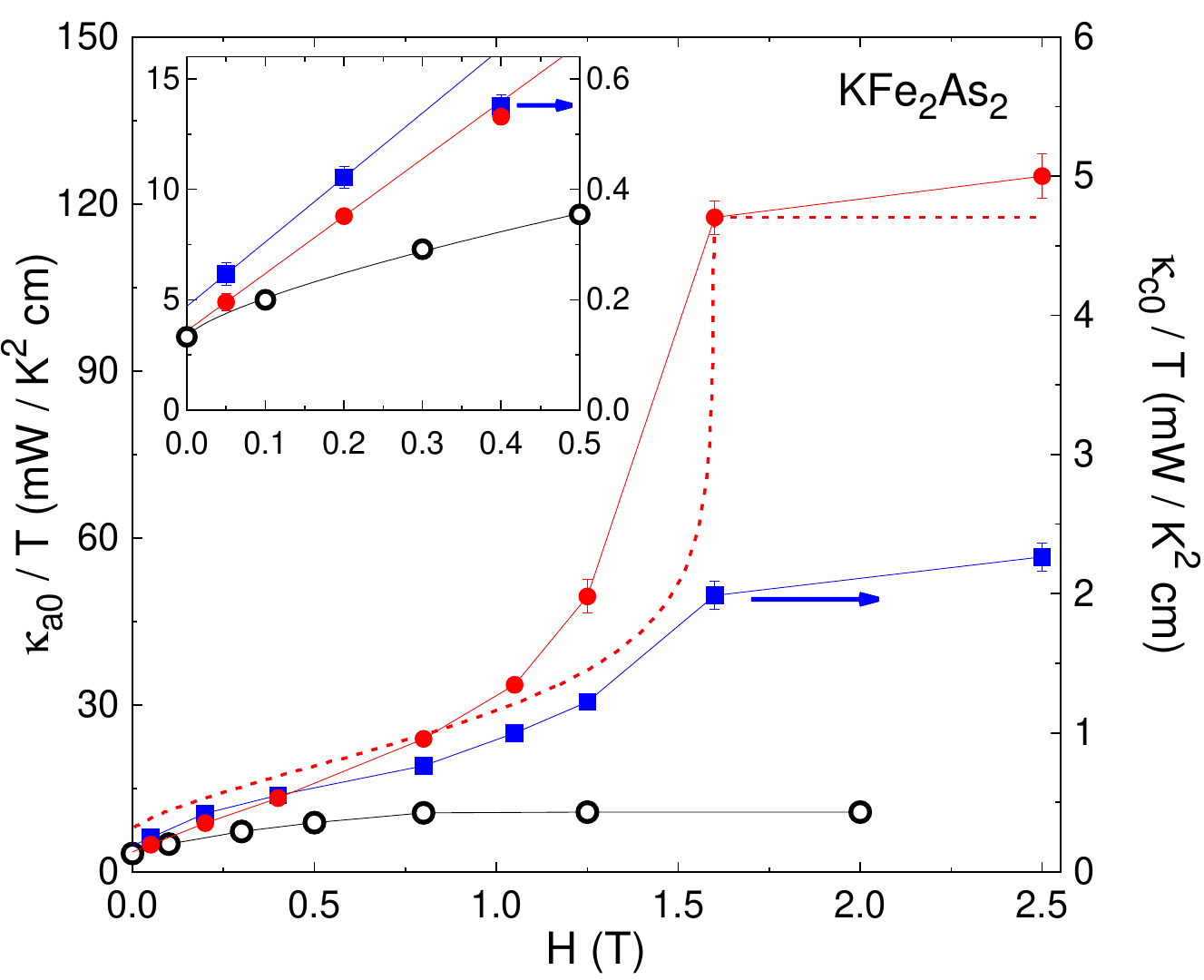}
\caption{
Field dependence of $\kappa_0/T$ obtained as in Fig.~2 (with corresponding symbols). 
The dashed line is a theoretical calculation for a $d$-wave superconductor with $\hbar \Gamma/\Delta_0 = 0.1$~\cite{Vekhter1999}. 
{\it Inset}:
Zoom at low field.
Lines are a power-law fit to extract the value of $\kappa_0/T$ at $H=0$.
}
\label{Fig3}
\end{figure}


To compare with cuprates, the archetypal $d$-wave superconductors,
we use Eq.~1 expressed directly in terms of $v_{\Delta}$, the slope of the gap at the node, namely
$\kappa_{00}/T \simeq (k_{\rm B}^2/3\hbar c) (v_{\rm F}/v_{\Delta})$, with $c$ the interlayer separation~\cite{Graf1996,Durst2000}.
The ratio $v_{\rm F}/v_{\Delta}$ was measured by ARPES on Ba$_2$Sr$_2$CaCu$_2$O$_{8+\delta}$~\cite{Vishik2010}, 
giving  $v_{\rm F}/v_{\Delta} \simeq 16$ at optimal doping, so that
$\kappa_{00}/T \simeq 0.16$~mW/K$^2$~cm.
This is in excellent agreement with the experimental
value of  $\kappa_0/T = 0.15 \pm 0.01$~mW/K$^2$~cm measured in YBa$_2$Cu$_3$O$_y$ at optimal doping~\cite{Hill2004}.

In Fig.~4a, we plot 
$\kappa_0/T$ vs $\Gamma$ for both KFe$_2$As$_2$ and YBa$_2$Cu$_3$O$_7$,
the superconductor in which universal heat transport was first demonstrated~\cite{Taillefer1997}.
We see that $\kappa_0/T$ remains approximately constant up to at least $\hbar \Gamma \simeq 0.5~k_{\rm B} T_{c0}$ in both cases.
 We conclude that both the magnitude of $\kappa_0/T$ in KFe$_2$As$_2$ and its insensitivity to impurity scattering are precisely those expected of
 a $d$-wave superconductor. 
 By contrast, in an extended $s$-wave superconductor, there is no direct relation between $\kappa_0/T$ and $\Delta_0$, and
 a strong non-monotonic dependence on $\Gamma$ is expected, since impurity scattering will inevitably make $\Delta_0$ less anisotropic~\cite{Borkowski1994}.
 This is confirmed by calculations applied to pnictides, which typically find that $\kappa_0/T$ vs $\Gamma$ first rises, and then plummets to zero
 when nodes are lifted by  strong scattering~\cite{Mishra2009} (see Fig. 4a).


{\it Critical scattering rate.--}
In a $d$-wave superconductor, the critical scattering rate $\Gamma_c$ is such that
$\hbar \Gamma_c \simeq k_{\rm B} T_{c0}$~\cite{Alloul2009}.
 We can estimate $\Gamma_c$ for KFe$_2$As$_2$ 
 from the critical value of $\rho_0$, evaluated as twice that for which $\Gamma/\Gamma_c = 0.5$ in
 Fig.~1c, namely $\rho_0^{\rm crit} \simeq 4.5~\mu\Omega$~cm.
 Using $L_0 / \rho_0^{\rm crit}  = \gamma_{\rm N} v_{\rm F}^2 \tau_c / 3$,
where $L_0 \equiv (\pi^2 / 3) (k_{\rm B}/e)^2$,   
we get
$\hbar \Gamma_c = \hbar /2\tau_c  \simeq 1.3 \pm 0.2~k_{\rm B} T_{c0}$,
in excellent agreement with expectation for a $d$-wave state.
 By contrast, 
  $\hbar \Gamma_c / k_{\rm B} T_{c0} \simeq 45$
 in BaFe$_2$As$_2$ and SrFe$_2$As$_2$ at optimal Co, Pt or Ru doping~\cite{Kirshenbaum2012} (see Fig.~4b).
 This factor 30 difference in the sensitivity of $T_c$ to impurity scattering is proof
 that the pairing symmetry of KFe$_2$As$_2$ is different from the $s$-wave symmetry
 of Co-doped BaFe$_2$As$_2$~\cite{Reid2010}.


\begin{figure}[t]
\centering
\includegraphics[width=8.5cm]{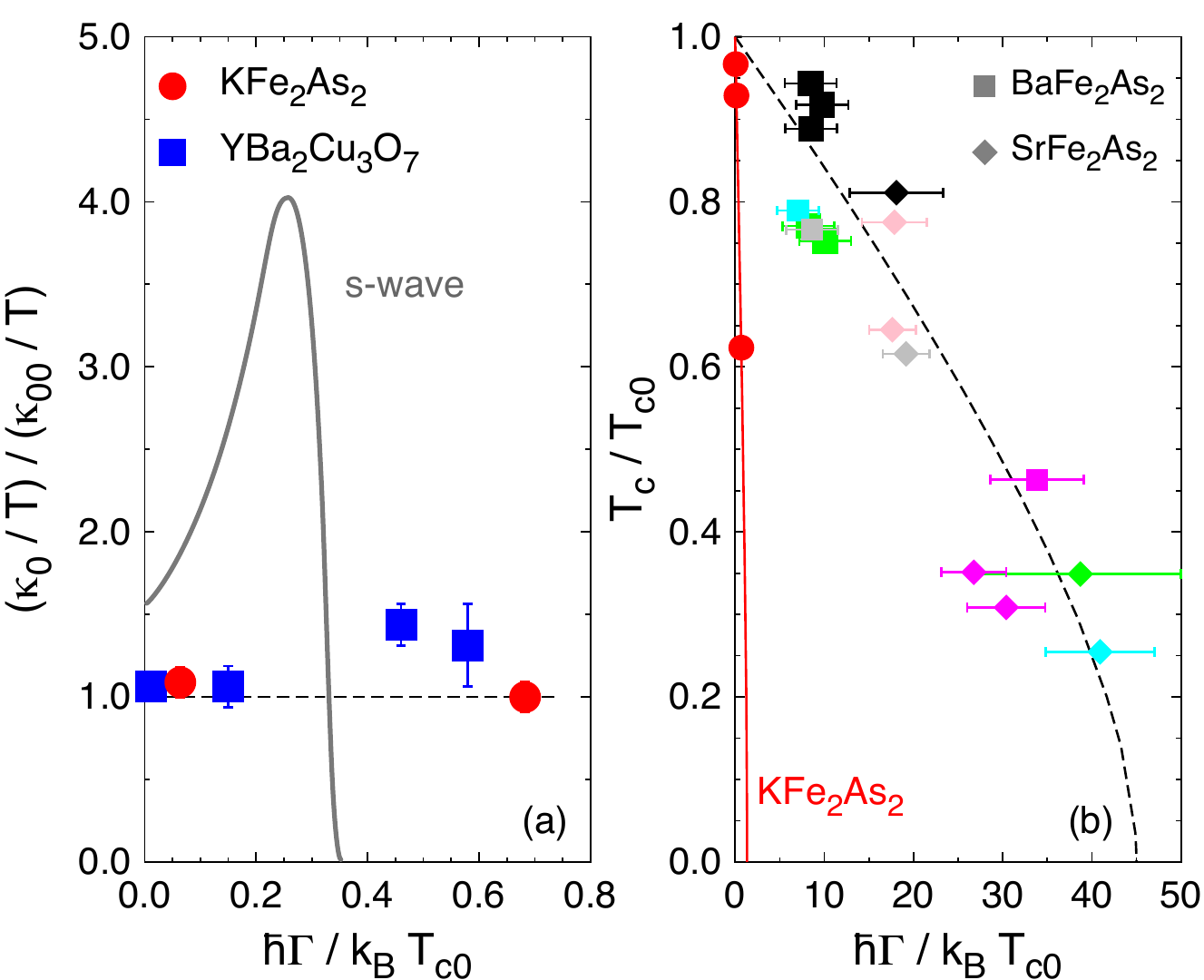}
\caption{
Dependence of $\kappa_0/T$ (a) and $T_c$ (b) on impurity  scattering rate $\Gamma$,
normalized by $T_{c0}$, the disorder-free superconducting temperature. 
(a) $\kappa_0/T$ for KFe$_2$As$_2$ (red circles; see text) and the cuprate YBa$_2$Cu$_3$O$_7$ (blue squares; from ref.~\onlinecite{Taillefer1997}),
normalized by the theoretically expected value for a $d$-wave superconductor, $\kappa_{00}/T = 3.3$ and 0.16 mW/K$^2$~cm, respectively (see text).
The typical dependence expected of an $s$-wave state with accidental nodes is also shown, 
from a calculation applied to pnictides (black line; from ref.~\onlinecite{Mishra2009}).
(b) $T_c$ for KFe$_2$As$_2$ (red circles; from Fig. 1c) and for the pnictides BaFe$_2$As$_2$ and SrFe$_2$As$_2$ at optimal doping
(from ref.~\onlinecite{Kirshenbaum2012}).
%
%
}
\label{Fig2b}
\end{figure}




{\it Direction dependence}.Ð In the case of a $d$-wave gap on a single quasi-2D cylindrical Fermi surface (at the zone center), the gap would necessarily have 4 line nodes that run ÓverticallyÓ along the $c$ axis. In such a nodal structure, zero-energy nodal quasiparticles will conduct heat not only in the plane, but also along the $c$ axis, by an amount proportional to the $c$-axis dispersion of the Fermi surface. In the simplest case, $c$-axis conduction will be smaller than $a$-axis conduction by a factor equal to the mass tensor anisotropy ($v_{F}^2$ in Eq.~1). In other words, ($\kappa_{a0}/T$)/($\kappa_{c0}/T$) $\simeq$ ($\kappa_{\rm aN}/T$)/($\kappa_{\rm cN}/T$) = ($\sigma_{\rm aN}$)/($\sigma_{\rm cN}$), the anisotropy in the normal-state thermal and electrical conductivities, respectively. This is confirmed by calculations for a quasi-2D $d$-wave superconductor \cite{Vekhter2007}, whose vertical line nodes yield an anisotropy ratio in the superconducting state very similar to that of the normal state. This is what we see in  KFe$_2$As$_2$ (inset of Fig.~3): ($\kappa_{a0}/T$)/($\kappa_{c0}/T$) = $20 \pm 4$, very close to the intrinsic normal-state anisotropy ($\sigma_{\rm aN}$)/($\sigma_{\rm cN}$) = ($\Delta \rho_c$)/($\Delta \rho_a$) = $25 \pm 1$. This shows that the nodes in KFe$_2$As$_2$ are vertical lines running along the $c$ axis, ruling out proposals \cite{Suzuki2011} of horizontal line nodes lying in a plane normal to the $c$ axis. 

Moreover, the fact that the Fermi surface of KFe$_2$As$_2$ contains several sheets with very different $c$-axis dispersions \cite{Terashima2010a,Yoshida2012} provides compelling evidence in favor of $d$-wave symmetry. 
%
%
In an extended $s$-wave scenario, the gap would typically develop vertical line nodes on some but not all zone-centered sheets of the Fermi surface \cite{Maiti2011}, and so the anisotropy in $\kappa$ would typically be very different in the superconducting and normal states, unlike what is measured. By contrast, in $d$-wave symmetry all zone-centered sheets must necessarily have nodes, thereby ensuring automatically that transport anisotropy remains similar in the superconducting and normal states.

{\it Temperature dependence.--}
So far, we have discussed the limit $T \to 0$ and $H \to 0$, where nodal quasiparticles are excited only by the pair-breaking effect of impurities.
Raising the temperature will further excite nodal quasiparticles.
Calculations for a $d$-wave superconductor show that the electronic thermal conductivity 
grows as $T^2$~\cite{Graf1996,Graf1996a}:
\begin{equation}
\frac{\kappa}{T} ~\simeq ~ \frac{\kappa_{00}}{T} ~( 1 + a \frac{T^2}{\gamma^2} )~~~~~~~~~,
\end{equation}
where $a$ is a dimensionless number and $\hbar \gamma$ is the impurity bandwidth,
which grows with the scattering rate $\Gamma$~\cite{Graf1996}.
A $T^2$ slope in $\kappa/T$ was resolved
in YBa$_2$Cu$_3$O$_7$~\cite{Hill2004}.

Our KFe$_2$As$_2$ sample shows a clear $T^2$ dependence below 
$T \simeq 0.3$~K, with 
$\kappa_{a}/T = (\kappa_{a0}/T) ( 1 + 23~T^2)$ (Fig.~2).
Comparison with the data by Dong {\it et al}.~\cite{Dong2010} reveals that 
this $T^2$ term must be due to quasiparticles.
Indeed, because phonon conduction at sub-Kelvin temperatures is limited by sample boundaries
and not impurities~\cite{Li2008}, the fact that the slope of $\kappa/T$ in their sample 
(of similar dimensions)
is at least 10 times smaller
(Fig.~2), 
implies that the larger slope in our data
must be electronic.

In the limit of unitary scattering, $\gamma^2 \propto \Gamma$, so that a 10-times larger $\Gamma$
would yield a 10-times smaller $T^2$ slope~\cite{Graf1996}, consistent with observation.
The temperature below which the $T^2$ dependence of $\kappa_e/T$ sets in, $T \simeq0.1~T_c$,
is a measure of $\gamma$.
It is in excellent agreement with the temperature below 
which the penetration depth $\lambda_a(T)$ of KFe$_2$As$_2$ (in a sample with similar RRR) 
deviates from its linear $T$ dependence~\cite{Hashimoto2010a}, 
as expected of a $d$-wave superconductor~\cite{Hirschfeld1993}.
%
Note that the $T$ dependence of $\kappa/T$ for an extended $s$-wave gap
is not $T^2$~\cite{Mishra2009}.


{\it Magnetic field dependence.--}
Increasing the magnetic field is another way to excite quasiparticles.
If the gap has nodes, 
the field will cause an immediate rise in $\kappa_0/T$~\cite{Shakeripour2009,Kubert1998,Vekhter1999},
as observed in all three samples of KFe$_2$As$_2$ (inset of Fig.~3).
Calculations for a $d$-wave superconductor in the clean limit ($\hbar \Gamma \ll k_{\rm B} T_c$)
yield a non-monotonic increase of $\kappa_0/T$ vs $H$~\cite{Vekhter1999}
%
in remarkable agreement with data on the clean sample (Fig.~3).

A rapid initial rise in $\kappa_0/T$ vs $H$ has been observed in the cuprate superconductors
 YBa$_2$Cu$_3$O$_7$~\cite{Chiao1999} and Tl$_2$Ba$_2$CuO$_{6+\delta}$~\cite{Proust2002}.
In the dirty limit, KFe$_2$As$_2$~\cite{Dong2010} and Tl$_2$Ba$_2$CuO$_{6+\delta}$~\cite{Proust2002} show nearly identical 
curves of $\kappa_0/T$ vs $H/H_{c2}$ (see ref.~\onlinecite{Dong2010}).
Measurements on cuprates in the clean limit, such as optimally-doped YBa$_2$Cu$_3$O$_y$, have so far been limited to
$H \ll H_{c2} $.
%


In summary, all aspects of the thermal conductivity of KFe$_2$As$_2$, including its dependence on impurity scattering, current direction,  
temperature and magnetic field, are in detailed and quantitative agreement with theoretical calculations for a $d$-wave superconductor.
The scattering rate needed to suppress $T_c$ to zero is exactly as expected of $d$-wave symmetry, and vastly smaller
than that found in other pnictide superconductors where the pairing symmetry is believed to be $s$-wave.
This is compelling evidence that the pairing symmetry in this iron-arsenide superconductor is $d$-wave, 
in agreement with renormalization-group calculations 
~\cite{Thomale2011}.
Replacing K in KFe$_2$As$_2$ by Ba leads to a superconducting state with a 10 times higher $T_c$, but with a full gap without nodes~\cite{Reid2011},
necessarily of a different symmetry.
Understanding the relation between this factor 10 and the pairing symmetry provides insight into what boosts $T_c$ in these systems.
%

We thank A.~Carrington, J. Chang, A.~Chubukov, R. Fernandes, R.~W.~Hill, P.~J.~Hirschfeld, J. Paglione, S.~Y.~Li, M.~Sutherland, R.~Thomale and I.~Vekhter for fruitful discussions and J. Corbin for his assistance with the experiments. 
%
Work at Sherbrooke was supported by a Canada Research Chair, CIFAR, NSERC, CFI and FQRNT.
Work at the Ames Laboratory was supported by the DOE-Basic Energy Sciences under Contract No. DE-AC02-07CH11358.
Work in Japan was supported by Grants-in-Aid for Scientific Research (Nos. 21540351 \& 22684016) 
from MEXT
and JSPS
and Innovative Areas ``Heavy Electrons" (Nos. 20102005 \& 21102505) from MEXT, 
Global COE and AGGST financial support program from Chiba University.

\end{document}